\documentclass[twocolumn]{aastex701}

\newcommand{\Msun}{\ensuremath{{M}_{\odot}}}

\newcommand{\Rsun}{\ensuremath{{\rm R_{\odot}}}}

\usepackage{amsmath}
\usepackage{tabularx}
\usepackage{multirow}
\usepackage{subcaption}
\usepackage{natbib}

\begin{document}

\newcommand{\ih}[1]{\textcolor{blue}{\textbf{IH:} #1}}
\newcommand{\ch}[1]{\textcolor{purple}{\textbf{CH:} #1}}
\newcommand{\cass}[1]{\textcolor{teal}{\textbf{Cass:} #1}}
\newcommand{\td}[1]{\textcolor{cyan}{\textbf{TD:} #1}}
\newcommand{\tom}[1]{\textcolor{magenta}{\textbf{TH:} #1}}
\newcommand{\cp}[1]{\textcolor{red}{\textbf{CP:} #1}}
\newcommand{\red}[1]{\textcolor{red}{#1}}
\newcommand*\edits[1]{\textcolor{purple}{\textbf{#1}}}

\title{exoALMA XIX: Confirmation of non-thermal line broadening in the DM Tau protoplanetary disk}

\correspondingauthor{Caitlyn Hardiman}
\email[show]{caitlyn.hardiman@monash.edu}



\author[0009-0003-7403-9207]{Caitlyn Hardiman}
\affiliation{School of Physics \& Astronomy, Monash University, Clayton VIC 3800, Australia}
\email{}

\author[0000-0001-5907-5179]{Christophe Pinte}
\affiliation{Univ. Grenoble Alpes, CNRS, IPAG, 38000 Grenoble, France}
\email{}

\author[0000-0002-4716-4235]{Daniel J. Price}    
\affiliation{School of Physics \& Astronomy, Monash University, Clayton VIC 3800, Australia}
\email{}

\author[0000-0001-7641-5235]{Thomas Hilder}
\affiliation{School of Physics \& Astronomy, Monash University, Clayton VIC 3800, Australia}
\email{}

\author[0000-0003-1502-4315]{Iain Hammond}
\affiliation{Max-Planck Institute for Astronomy (MPIA), Königstuhl 17, 69117 Heidelberg, Germany}
\affiliation{School of Physics \& Astronomy, Monash University, Clayton VIC 3800, Australia}
\email{}

\author[0000-0002-1283-6038]{Ta\"issa Danilovich}
\affiliation{School of Physics \& Astronomy, Monash University, Clayton VIC 3800, Australia}
\affiliation{Institute of Astronomy, KU Leuven, Celestijnenlaan 200D,  3001 Leuven, Belgium}
\email{}

\author[0000-0003-2253-2270]{Sean M. Andrews}
\affiliation{Center for Astrophysics | Harvard \& Smithsonian, Cambridge, MA 02138, USA}
\email{}

\author[0000-0003-1534-5186]{Richard Teague}
\affiliation{Department of Earth, Atmospheric, and Planetary Sciences, Massachusetts Institute of Technology, Cambridge, MA 02139, USA}
\email{}

\author[0000-0003-4853-5736]{Giovanni Rosotti}
\affiliation{Dipartimento di Fisica, Universit\`a degli Studi di Milano, Via Celoria 16, 20133 Milano, Italy}
\email{}

\author[0000-0002-9298-3029]{Mario Flock}
\affiliation{Max-Planck Institute for Astronomy (MPIA), Königstuhl 17, 69117 Heidelberg, Germany}
\email{}

\author[0000-0002-2700-9676]{Gianni Cataldi}
\affiliation{National Astronomical Observatory of Japan, 2-21-1 Osawa, Mitaka, Tokyo 181-8588, Japan}
\email{}


\author[0000-0001-7258-770X]{Jaehan Bae}
\affiliation{Department of Astronomy, University of Florida, Gainesville, FL 32611, USA}
\email{}

\author[0000-0001-6378-7873]{Marcelo Barraza-Alfaro}
\affiliation{Department of Earth, Atmospheric, and Planetary Sciences, Massachusetts Institute of Technology, Cambridge, MA 02139, USA}
\email{}

\author[0000-0002-7695-7605]{Myriam Benisty}
\affiliation{Université Côte d'Azur, Observatoire de la Côte d'Azur, CNRS, Laboratoire Lagrange, France}
\affiliation{Max-Planck Institute for Astronomy (MPIA), Königstuhl 17, 69117 Heidelberg, Germany}
\email{}

\author[0000-0003-3713-8073]{Nicolás Cuello}
\affiliation{Univ. Grenoble Alpes, CNRS, IPAG, 38000 Grenoble, France}
\email{}

\author[0000-0003-2045-2154]{Pietro Curone}
\affiliation{Departamento de Astronomía, Universidad de Chile, Camino El Observatorio 1515, Las Condes, Santiago, Chile}
\email{}

\author[0000-0002-1483-8811]{Ian Czekala}
\affiliation{School of Physics \& Astronomy, University of St. Andrews, North Haugh, St. Andrews KY16 9SS, UK}
\email{}

\author[0000-0003-4689-2684]{Stefano Facchini}
\affiliation{Dipartimento di Fisica, Università degli Studi di Milano, Via Celoria 16, I-20133 Milano, Italy}
\email{}

\author[0000-0003-4679-4072]{Daniele Fasano}
\affiliation{Université Côte d'Azur, Observatoire de la Côte d'Azur, CNRS, Laboratoire Lagrange, France}
\affiliation{Max-Planck Institute for Astronomy (MPIA), Königstuhl 17, 69117 Heidelberg, Germany}
\email{}

\author[0000-0003-1117-9213]{Misato Fukagawa}
\affiliation{National Astronomical Observatory of Japan, 2-21-1 Osawa, Mitaka, Tokyo 181-8588, Japan}
\email{}

\author[0000-0002-5503-5476]{Maria Galloway-Sprietsma}
\affiliation{Department of Astronomy, University of Florida, Gainesville, FL 32611, USA}
\email{}

\author[0000-0002-5910-4598]{Himanshi Garg}
\affiliation{School of Physics \& Astronomy, Monash University, Clayton VIC 3800, Australia}
\email{}

\author[0000-0002-8138-0425]{Cassandra Hall}
\affiliation{Department of Physics and Astronomy, The University of Georgia, Athens, GA 30602, USA}
\affiliation{Center for Simulational Physics, The University of Georgia, Athens, GA 30602, USA}
\affiliation{Institute for Artificial Intelligence, The University of Georgia, Athens, GA, 30602, USA}
\email{}

\author[0000-0001-6947-6072]{Jane Huang}
\affiliation{Department of Astronomy, Columbia University, 538 W. 120th Street, Pupin Hall, New York, NY 10027, USA}
\email{}

\author[0000-0003-1008-1142]{John D. Ilee}
\affiliation{School of Physics and Astronomy, University of Leeds, Leeds, UK, LS2 9JT}
\email{}

\author[0000-0001-8446-3026]{Andres F. Izquierdo}
\altaffiliation{NASA Hubble Fellowship Program Sagan Fellow}
\affiliation{Department of Astronomy, University of Florida, Gainesville, FL 32611, USA}
\affiliation{Leiden Observatory, Leiden University, P.O. Box 9513, NL-2300 RA Leiden, The Netherlands}
\affiliation{European Southern Observatory, Karl-Schwarzschild-Str. 2, D-85748 Garching bei München, Germany}
\email{}

\author[0000-0001-7235-2417]{Kazuhiro Kanagawa}
\affiliation{College of Science, Ibaraki University, 2-1-1 Bunkyo, Mito, Ibaraki 310-8512, Japan}
\email{}

\author[0000-0002-8896-9435]{Geoffroy Lesur}
\affiliation{Univ. Grenoble Alpes, CNRS, IPAG, 38000 Grenoble, France}
\email{}

\author[0000-0002-2357-7692]{Giuseppe Lodato}
\affiliation{Dipartimento di Fisica, Universit\`a degli Studi di Milano, Via Celoria 16, 20133 Milano, Italy}
\email{}

\author[0000-0003-4663-0318]{Cristiano Longarini}
\affiliation{Institute of Astronomy, University of Cambridge, Madingley Rd, CB30HA, Cambridge, UK}
\affiliation{Dipartimento di Fisica, Universit\`a degli Studi di Milano, Via Celoria 16, 20133 Milano, Italy}
\email{}

\author[0000-0002-8932-1219]{Ryan Loomis}
\affiliation{National Radio Astronomy Observatory, Charlottesville, VA 22903, USA}
\email{}

\author[0000-0002-1637-7393]{Francois Menard}
\affiliation{Univ. Grenoble Alpes, CNRS, IPAG, 38000 Grenoble, France}
\email{}

\author[0000-0003-4039-8933]{Ryuta Orihara}
\affiliation{College of Science, Ibaraki University, 2-1-1 Bunkyo, Mito, Ibaraki 310-8512, Japan}
\email{}

\author[0000-0002-0491-143X]{Jochen Stadler}
\affiliation{Université Côte d'Azur, Observatoire de la Côte d'Azur, CNRS, Laboratoire Lagrange, France}
\email{}

\author[0000-0003-1412-893X]{Hsi-Wei Yen}
\affiliation{Academia Sinica Institute of Astronomy \& Astrophysics, 11F of Astronomy-Mathematics Building, AS/NTU, No.1, Sec. 4, Roosevelt Rd, Taipei 10617, Taiwan}
\email{}

\author[0000-0002-3468-9577]{Gaylor Wafflard-Fernandez}
\affiliation{Univ. Grenoble Alpes, CNRS, IPAG, 38000 Grenoble, France}
\email{}

\author[0000-0003-1526-7587]{David J. Wilner}
\affiliation{Center for Astrophysics | Harvard \& Smithsonian, Cambridge, MA 02138, USA}
\email{}

\author[0000-0002-7501-9801]{Andrew J. Winter}
\affiliation{Astronomy Unit, School of Physics and Astronomy, Queen Mary University of London, London E1 4NS, UK}
\email{}

\author[0000-0002-7212-2416]{Lisa W\"olfer}
\affiliation{Department of Earth, Atmospheric, and Planetary Sciences, Massachusetts Institute of Technology, Cambridge, MA 02139, USA}
\email{}

\author[0000-0001-8002-8473]{Tomohiro C. Yoshida}
\affiliation{National Astronomical Observatory of Japan, 2-21-1 Osawa, Mitaka, Tokyo 181-8588, Japan}
\email{}

\author[0000-0001-9319-1296]{Brianna Zawadzki}
\affiliation{Department of Astronomy, Van Vleck Observatory, Wesleyan University, 96 Foss Hill Drive, Middletown, CT 06459, USA}
\affiliation{Department of Astronomy \& Astrophysics, 525 Davey Laboratory, The Pennsylvania State University, University Park, PA 16802, USA}
\email{}

\begin{abstract}

Turbulence is expected to transport angular momentum and drive mass accretion in protoplanetary disks. One way to directly measure turbulent motion in disks is through molecular line broadening. 
DM~Tau is one of only a few disks with claimed detection of non-thermal line broadening of 0.25--0.33~$c_{\rm s}$, where $c_{\rm s}$ is the sound speed.
Using the radiative transfer code {\sc mcfost} within a Bayesian inference framework that evaluates over five million disk models to efficiently sample the parameter space, we fit high-resolution ($0\farcs15$, 28~m~s$^{-1}$) $^{12}$CO $J = 3 - 2$ observations of DM~Tau from the exoALMA Large Program.  
This approach enables us to simultaneously constrain the disk structure and kinematics, revealing a significant non-thermal contribution to the line width of $\sim$0.4~$c_{\rm s}$, inconsistent with purely thermal motions. 
Using the CO-based disk structure as a starting point, we reproduce the CS $J = 7 - 6$ emission well, demonstrating that the CS (which is more sensitive to non-thermal motions than CO) agrees with the turbulence inferred from the CO fit.
Establishing a well-constrained background disk model further allows us to identify residual structures in the moment maps that deviate from the expected emission, revealing localized perturbations that may trace forming planets. 
This framework provides a powerful general approach for extracting disk structure and non-thermal broadening directly from molecular line data, and can be applied to other disks with high-quality observations.

\end{abstract}


\section{Introduction} \label{sec:intro}

Turbulent motions are one of the main contenders for driving angular momentum transport in protoplanetary disks, with $\alpha$-disk models \citep{shakura73, pringle81} parameterising the bulk disk viscosity $\nu$ in terms of the dimensionless parameter $\alpha$ via
\begin{equation}
    \nu = \alpha c_{\rm s} H,
\end{equation}
where $c_{\rm s}$ is the sound speed and $H$ the pressure scale height. This relation assumes that the largest scale motions would be less than or equal to the scale height of the disk, $H$, and typical motions would be subsonic, implying $\alpha~\in~[0,1]$. The actual value of $\alpha$, and the source of the turbulence itself, are poorly constrained. 

High-spectral and spatial-resolution surveys such as exoALMA \citep{teague_exoALMA} now enable direct comparisons between observations and theoretical models of turbulence in protoplanetary disks \citep{barraza-alfaro_exoALMA}. One of the most prominent proposed mechanisms is the magneto-rotational instability \citep[MRI;][]{balbus91, balbus98}. 
Other candidates include gravitational instability \citep[GI, see][for a review]{kratter16}, embedded planets \citep{goodman01}, the vertical shear instability \citep[VSI, ][]{nelson13, stoll14, flock17} the convective overstability and the subcritical baroclinic instability \citep{pfeil2019}. Determining which of these mechanisms actually operate in disks requires observational constraints on the strength and spatial distribution of turbulence.

We can \emph{indirectly} measure $\alpha$ for protoplanetary disks via observations of accretion rates onto their host stars, which \citet{manara16} have found to span a range of $10^{-11} - 10^{-7} \Msun \ \rm yr^{-1}$. This can then be related to an $\alpha$ as $\dot{M}_{\rm acc} = 3\pi \nu \Sigma$, either assuming or fitting simultaneously for a surface density profile $\Sigma$ \citep[e.g.,][]{vandermarel21}. Similarly, one can match the observed disk lifetime to that predicted from self-similar disk evolution given values of $M_{\rm disk}$ and $\dot{M}_{\rm acc}$ \citep{hartmann98,manara23}. Based on these approaches, estimates of $\alpha$ typically fall in the range $10^{-4} - 10^{-3}$ \citep[see review by][]{rosotti23}. However, it is important to note that these measurements are instantaneous values that may vary significantly with time, particularly if the disk undergoes episodic accretion events such as FU~Ori-type outbursts.

Another indirect method of constraining turbulence in disks is by observing the height of the dust layer in edge-on disks, as turbulence will stir the dust up from the disk midplane \citep[e.g.][]{pinte16, villenave25}. Dust settling prescriptions \citep{dubrulle95, fromang09} relate the scale height of the dust to the gas scale height via a combination of an $\alpha$ and the stopping time of the dust grains where $H \propto \sqrt{\alpha/\tau_{\rm s}}$. Infrared and mm-imaging of edge-on disks clearly show that grains larger than $\approx~100 \ \mu\rm m$ are settled in the midplane of the disk \citep{villenave20, villenave22, villenave24, duchene24}, with the dust scale height up to an order of magnitude smaller than that of the gas, again indicating weak stirring of the dust layer, implying $\alpha \lesssim 10^{-4}$.

The first attempt to directly measure non-thermal molecular line broadening in protoplanetary disks was made by \citet{hughes11} using pre-ALMA Submillimeter Array observations of the TW~Hya and HD~163296 disks, which found an upper limit of $\lesssim$0.1$c_{\rm s}$ and a tentative detection of $\sim$0.4$c_{\rm s}$, respectively. With the advent of ALMA, higher spatial and spectral resolution data enabled much tighter constraints: \citet{flaherty15,flaherty17} reported $v_{\rm turb} \lesssim 0.03$--$0.06 c_{\rm s}$ in HD~163296; \citet{teague16} between 0.2--0.4~$c_{\rm s}$ in TW~Hya, revised downwards to $\lesssim 0.08 c_{\rm s}$ ($\alpha \lesssim 0.007$) by \citet{flaherty18}. Most recently, \citet{flaherty24} and \citet{panequecarreno24} reported 0.18--0.3$c_{\rm s}$ and 0.4--0.6$c_{\rm s}$, respectively, around IM~Lup.

As one of these few disks with a measured non-zero turbulence, DM~Tau offers the opportunity to probe non-thermal line broadening using the high resolution data from exoALMA. 
DM~Tau is a low mass ($\approx 0.5 \ M_\odot$) T-Tauri star with a large ($\approx 300$ au) protoplanetary disk featuring an inner disk and three pairs of gaps and rings seen in continuum emission \citep{andrews11, kudo18, francis22, curone_exoALMA}.  Its accretion rate has been measured as $\dot{M}_{\rm acc}=10^{-8.2} \ M_\odot \ \rm yr^{-1}$ \citep{manara14}. Line-broadening measurements by \citet{flaherty20} found 0.25--0.33~$c_{\rm s}$ (equivalent to 0.06--0.11 in $\alpha$), using a fixed stellar mass of 0.54 $\Msun$. \citet{guilloteau12} obtained consistent results using CS, a heavier molecule less sensitive to thermal broadening than CO, finding a turbulent linewidth of $\sim0.14$~km~s$^{-1}$ at 300~au. However, this is discrepant with the inferred value of $\alpha$ needed to explain the continuum gap depth at 21 au if carved by a planet ($\alpha \lesssim 10^{-3}$; \citealt{francis22}); we note, however, that the former measurements probe the outer disk while the latter constraint applies to the inner regions, and these estimates are therefore unlikely to correspond to the same radial location.

When modelling spatially resolved disks, accurately constraining non-thermal line broadening requires distinguishing its contribution from other factors that influence the observed kinematics, such as the stellar mass and disk temperature structure. The high angular and spectral resolution of the exoALMA Large Program \citep{teague_exoALMA} provides the level of detail needed to disentangle these effects. We use this data to quantify the level of non-thermal broadening in DM~Tau and to test whether a disk structure inferred from CO can also reproduce the emission from heavier molecules such as CS.

The paper is organized as follows: Section \ref{sec:methods} describes the data, the modeling approach, and the image-plane fitting methodology. Section \ref{sec:results} presents the inferred non-thermal line broadening from CO and the CS modeling results. In Section \ref{sec:discussion}, we interpret these results in the context of disk kinematics and potential planet–disk interactions, and we summarize our findings in Section \ref{sec:conclusions}.

\section{Methods}\label{sec:methods}

\subsection{Observations}\label{sec:data}

We used $^{12}$CO \textit{J}=3--2 and CS \textit{J}=7--6 observations of DM~Tau from the exoALMA Large Program, with the data reduction detailed in \citet{loomis_exoALMA}. For CO, we selected a primary beam–corrected cube with a spatial resolution of 0\farcs15 and a spectral resolution of 28~m~s$^{-1}$. To reduce computational cost, we fit only 17 evenly-spaced representative velocity channels spanning 3.5 to 8.5~km~s$^{-1}$ around the systemic velocity \citep[$v_\mathrm{syst} = 6.03$~km~s$^{-1}$,][]{izquierdo_exoALMA}, which capture the bulk of the emission. We restricted ourselves to a 15\farcs0 box centered on the disk to encompass the full CO-emitting region. The CS cube used was also primary beam corrected, at the same spatial resolution but with a spectral resolution of 100~m~s$^{-1}$, needed to obtain sufficient signal-to-noise for this comparatively faint line. A 6\farcs0 box was used to capture the full CS emission while limiting data volume.

\subsection{Radiative transfer modeling}\label{sec:model}

We used the radiative transfer code \textsc{mcfost} \citep{pinte06, pinte09} to generate disk models of the $^{12}$CO and CS exoALMA data for DM~Tau, assuming an unperturbed smooth disk containing no planets or gaps. Both the models and the data include continuum emission, which we did not subtract \citep[as in previous works,][]{guilloteau12, flaherty20}, since the continuum is much more compact than the $^{12}$CO gas (119~au vs.\ 580~au enclosing 68$\%$ of the flux; \citealt{curone_exoALMA, longarini_exoALMA}) and only affects the innermost regions of the disk.

We used a tapered-edge disk model with the disk surface density profile $\Sigma$ set according to the self-similar solution of \citet{lynden-bell74} \citep[see e.g.][]{hartmann98, manara23} given by

\begin{equation}
    \Sigma(r) = \frac{M_0 (2-\gamma)}{2\pi R_{\rm c}^2} \left(\frac{r}{R_{\rm c}}\right)^{-\gamma} \exp\left[ - \left(\frac{r}{R_c}\right)^{2 - \gamma}\right],
\end{equation}

which exponentially decreases for radii $r$ greater than the critical radius $R_{\rm c}$. $\gamma$ is the power law index of the surface density and of the outer taper, and $M_0$ is the disk mass. We let our disk extend from an inner radius $R_{\rm in}$ of 5~au to $R_{\rm out}$ of 1000~au, noting that the outer edge is more properly constrained by $R_{\rm c}$, or the surface density gradient itself as $\gamma$ approaches $2$. We also assumed a Gaussian vertical density profile parameterized by a scale height with a radial dependence given by $h(r) = h_0 (r/r_0)^\Psi$ at a reference radius of $r_0 = 150$~au where $\Psi$ is the flaring exponent. 

The disk temperature structure was self-consistently computed by \textsc{mcfost} assuming local thermodynamic equilibrium (LTE). This temperature distribution, combined with the density structure, determines the regions affected by molecular processes. We included molecular photo-dissociation and photo-desorption, and imposed CO freeze-out in regions where the temperature drops below 20~K, using a depletion factor $f_{\rm remain}$ that sets the fraction of gas remaining in the gas phase \citep[following][]{pinte18photo}. By default, $f_{\rm remain} = 0$, meaning all CO in these regions is frozen out. For CS, we adopted a freeze-out temperature of 30~K, near the lower end of the range of values reported in the literature (20–30~K from excitation/evaporation estimates \citep{guilloteau12, legal19}, 50–60~K from other observational studies \citep{garrod06, vanderplas14}).

The external UV field was modeled as the \citet{draine78} interstellar radiation field scaled by a factor $\chi_{\rm ISM}$, which contributes extra heating in the disk's upper layers.

Vertical dust settling was implemented using the prescription given in Equation 19 of \citet{fromang09}, where the amount of dust settling is controlled by $\alpha_{\rm dust}$, and the Schmidt number in their Equation 20 was set to a fixed value of $S_{\rm c} = 1.5$. In \textsc{mcfost}, $\alpha_{\rm dust}$ only controls the diffusion coefficient for the dust and does not contribute to additional line broadening, which is implemented as an independent parameter (described in Section \ref{sec: fturb}).

We employed a dust population in our models from the DiscAnalysis (DIANA) project \citep{woitke16} for DM~Tau, which consisted of 79\% silicates \citep{dorschner95} and 21\% amorphous carbon \citep{zubko96}, with 25\% grain porosity. Following their setup, dust properties were calculated using the distribution of hollow spheres formalism \citep{min05} with a maximum void fraction of 0.8, and the grain size distribution was set as d$N(a) \propto a^{-3.7}$ d$a$ for grain sizes $a$ between $a_{\rm min} = 0.0152~\mu$m and $a_{\rm max} = 3990~\mu$m. We also used the DIANA stellar spectrum file and corresponding effective stellar temperature of 3779~K\footnote{The stellar spectrum file and the \textsc{MCFOST} parameter file used for the DIANA model are found at \href{https://doi.org/10.26180/30908507}{doi:10.26180/30908507}.}.

In addition to the parameters already described, we allowed the disk orientation (inclination $i$ and position angle PA), stellar properties (stellar mass $M_{\star}$ and radius $R_{\star}$, the latter of which allows the stellar luminosity to vary), and gas properties (gas mass $M_{\rm gas}$, gas-to-dust mass ratio, and CO abundance $X_{\rm CO}$) to vary when fitting the CO emission. Once the CO model was determined, we used it as the baseline for modeling CS, keeping the temperature and density distributions fixed and only fitting for the molecular abundance $X_{\rm CS}$ and the fraction of gas-phase CS in the freeze-out region $f_{\rm remain}$. We assumed constant abundances for both CO and CS throughout the disk, i.e., the molecular number density at each location is given by the product of the gas density and a fixed abundance, modulated by freeze-out, photo-dissociation, and photo-desorption where relevant.

Table \ref{tab:priors_posteriors} lists all of the parameters we allowed to vary for both the CO and CS fits, including the line broadening parameter $f_{\rm turb}$ we will describe more fully in Section \ref{sec: fturb}. We aimed to allow as much freedom in the model parameters as possible to avoid any degeneracies going unnoticed. Importantly, we included stellar mass as a variable parameter rather than using a literature value, which was the approach of \citet{flaherty20} and \citet{francis22}. In addition to using the discussed values from the DIANA project, we used a fixed distance to DM~Tau of 144.0 pc from \textit{Gaia} DR3 \citep{Gaia-Collaboration:2023} and a systemic velocity of 6.03~km~s$^{-1}$ \citep{izquierdo_exoALMA}.

\begin{table*}
\centering
\caption{Model parameters, priors, bounds, and posterior results for our fits of the DM Tau disk. CO emission and CS emission parameters are shown in separate blocks. For the CS fit, all other disk parameters were fixed to the CO best-fit values. Posterior values are the medians with 2$\sigma$ uncertainties.}
\label{tab:priors_posteriors}
\small 
\begin{tabular}{l | c | c | c}
\hline\hline
\multicolumn{4}{c}{\textbf{CO emission fit}} \\
\hline
Parameter & Prior & Bounds & Posterior Median \\
\hline
$i$ [$^\circ$] & $\mathcal{N}(39, 3)$ & (0, 180) & $39.7^{+0.7}_{-0.7}$ \\
$M_{\star}$ [$\Msun$] & $\mathcal{N}(0.45, 0.05)$ & (0, $\infty$) & $0.483^{+0.01}_{-0.01}$ \\
$h_0$ [au] & $\mathcal{N}(10, 2.5)$ & (0, $\infty$) & $20.6^{+0.6}_{-0.9}$ \\
$R_{\rm c}$ [au] & $\mathcal{N}(240, 20)$ & (7, $\infty$) & $336.4^{+27.0}_{-44.9}$ \\
$\Psi$ & $\mathcal{N}(1.315, 0.05)$ & (1, 2) & $1.16^{+0.023}_{-0.036}$ \\
PA [$^\circ$] & $\mathcal{N}(155.7, 3)$ & (0, 360) & $155.4^{+0.2}_{-0.2}$ \\
$\log_{10}(\alpha_{\rm dust})$ & $\mathcal{N}(-4, 0.5)$ & (-6, 0) & $-0.87^{+0.35}_{-0.20}$ \\
$f_{\rm turb}$ & $\mathcal{U}(0, 1)$ & (0.0, 1.0) & $0.404^{+0.016}_{-0.032}$ \\
$M_{\rm gas}$ [$\Msun$] & $\mathcal{N}(4\times10^{-2}, 1\times10^{-2})$ & (0, $\infty$) & $6.6^{+0.1}_{-0.1} \times 10^{-2}$ \\
$\log_{10}(M_{\rm gas}/M_{\rm dust})$ & $\mathcal{N}(2.0, 0.1)$ & (0, 3) & $2.50^{+0.19}_{-0.17}$ \\
$\log_{10}(X_{\rm CO})$ & $\mathcal{N}(-4.7, 1.5)$ & (-7, -2) & $-4.52^{+0.14}_{-0.07}$ \\
$f_{\rm remain}$ & $\mathcal{E}(0.2)$ & (0, 1) & $0.002^{+0.001}_{-0.001}$ \\
$\gamma$ & $\mathcal{N}(1.0, 0.1)$ & (0, 2) & $1.55^{+0.20}_{-0.17}$ \\
$R_{\star}$ [$\Rsun$] & $\mathcal{N}(1.22, 0.1)$ & (1, $\infty$) & $1.25^{+0.15}_{-0.13}$ \\
$\chi_{\rm ISM}$ & $\mathcal{E}(0.2)$ & — & $0.65^{+0.08}_{-0.10}$ \\
$\ln{q}$ & $\mathcal{N}(-1, 0.05)$ & (-3, 1) & $-1.72^{+0.08}_{-0.04}$ \\
\hline
\multicolumn{4}{c}{\textbf{CS emission fit (other disk parameters fixed)}} \\
\hline
Parameter & Prior & Bounds & Posterior Median \\
\hline
$\log_{10}(X_{\rm CS})$ & $\mathcal{N}(-8, 1.5)$ & (-12, -2) & $-9.4126^{+0.007}_{-0.006}$\\
$f_{\rm remain}$ & $\mathcal{E}(0.2)$ & (0, 1) & $0.997^{+0.003}_{-0.012}$\\
$\ln{q}$ & $\mathcal{N}(-1, 0.05)$ & (-3, 1) & $-1.22^{+0.01}_{-0.01}$\\
\hline
\end{tabular}
\normalsize
\vspace{0.5em}

\begin{flushleft}
$\mathcal{N}(\mu, \sigma)$: normal prior with mean $\mu$ and standard deviation $\sigma$ (truncated to bounds). \\
$\mathcal{U}(a, b)$: uniform prior between $a$ and $b$. \\
$\mathcal{E}(\lambda)$: exponential prior with scale $\lambda$ (truncated to bounds). \\
\end{flushleft}
\end{table*}

\subsection{Line broadening parameter}\label{sec: fturb}
Among the free parameters listed in Table~\ref{tab:priors_posteriors} is the non-thermal line broadening factor, $f_{\rm turb}$, which parametrizes the level of turbulent motion in the disk. We assume a Gaussian line profile of width

\begin{equation}\label{eq: global turb}
    \Delta V = \sqrt{\frac{2k_{\rm B}T(r,z)}{m_{\rm CO}} + v_{\rm turb}^2},
\end{equation}

where $k_{\rm B}$ is the Boltzmann constant, $T$ is the local kinetic temperature, $m_{\rm CO}$ is the mass of the CO molecule, and $v_{\rm turb}$ is the non-thermal line broadening. $\Delta V$ is $\sqrt{2} \sigma$ where $\sigma$ is the Gaussian standard deviation. Our prescription for $v_{\rm turb}$ follows theoretical predictions \citep{simon15,flock17} where the turbulence scales with the sound speed

\begin{equation}\label{eq: cs turb}
v_{\rm turb} = f_{\rm turb} \times  \sqrt{\frac{k_{\rm B}T(r, z)}{\mu m_{h}}},
\end{equation}

where $f_{\rm turb}$ is a constant and $\mu = 2.3$ is the mean molecular weight in units of the proton mass $m_{h}$.

\subsection{Parameter Estimation}

The problem with trying to measure non-thermal line broadening in a disk is that $v_{\rm turb}$ may be biased by other parameters in the fit, especially the thermal structure.
To account for this, we sampled the posterior distribution of the model parameters using a Markov Chain Monte Carlo (MCMC) framework, employing the affine-invariant ensemble sampler from the \textsc{emcee}\footnote{\url{https://github.com/dfm/emcee}} package \citep{emcee}. At each MCMC iteration, a synthetic data cube was produced with \textsc{mcfost}, convolved with the observational beam, and compared to the 17 equivalent velocity channels from the observed data cube selected to span the full disk emission (see Section~\ref{sec:data}).

To account for the possibility that the observational noise is underestimated, we included an additional free parameter, $\ln q$, which inflates the variance in each channel. In reality, uncertainties are underestimated because we neglect the stationary covariance between neighboring pixels \citep[e.g.][]{hilder_exoALMA}; including this parameter is intended to partially compensate for this effect.
With this term, the total variance is given by

\begin{equation}
    \mathbf{s_n}^2 = \sigma_n^2 + q^2\mathbf{\mathcal M}^2,
\end{equation}

where $\mathbf{\mathcal M}$ denotes the model flux at each pixel in the convolved channel and $\sigma_n$ is the measured RMS of the data cube. In addition to compensating for underestimated pixel-to-pixel uncertainties, $\ln q$ partially absorbs systematic mismatches in the absolute flux scale that may arise from amplitude calibration uncertainties. Since $q$ scales with the model flux, over- or under-estimates of the total flux are partially accounted for through this term, reducing their impact on the inferred physical parameters.
The parameter $\ln q$ is sampled alongside all other model parameters. 

Assuming Gaussian noise, the log-likelihood is then computed as

\begin{equation}
    \ln \mathcal{L} = -\frac{1}{2} \sum \left[ \frac{(\mathcal D - \mathcal M)^2}{s_n^2} + \ln(2\pi s_n^2) \right],
\end{equation}

with $\mathcal D$ representing the data values in the corresponding channels.

All fits used 512 walkers. The CO and CS fits were run for a minimum of 10000 and 6000 iterations respectively, discarding the first 4000 as burnin. For the main CO fit this therefore involved evaluating 5,120,000 radiative transfer models in 3D. Each model computation completed in around 4 minutes on a single CPU, in total requiring 55 days of wall time running in parallel on 256 CPUs.

\section{Results}\label{sec:results}

\subsection{CO fitting}\label{CO results} \label{sec:co results}

\begin{figure*}
    \includegraphics[width=\textwidth]{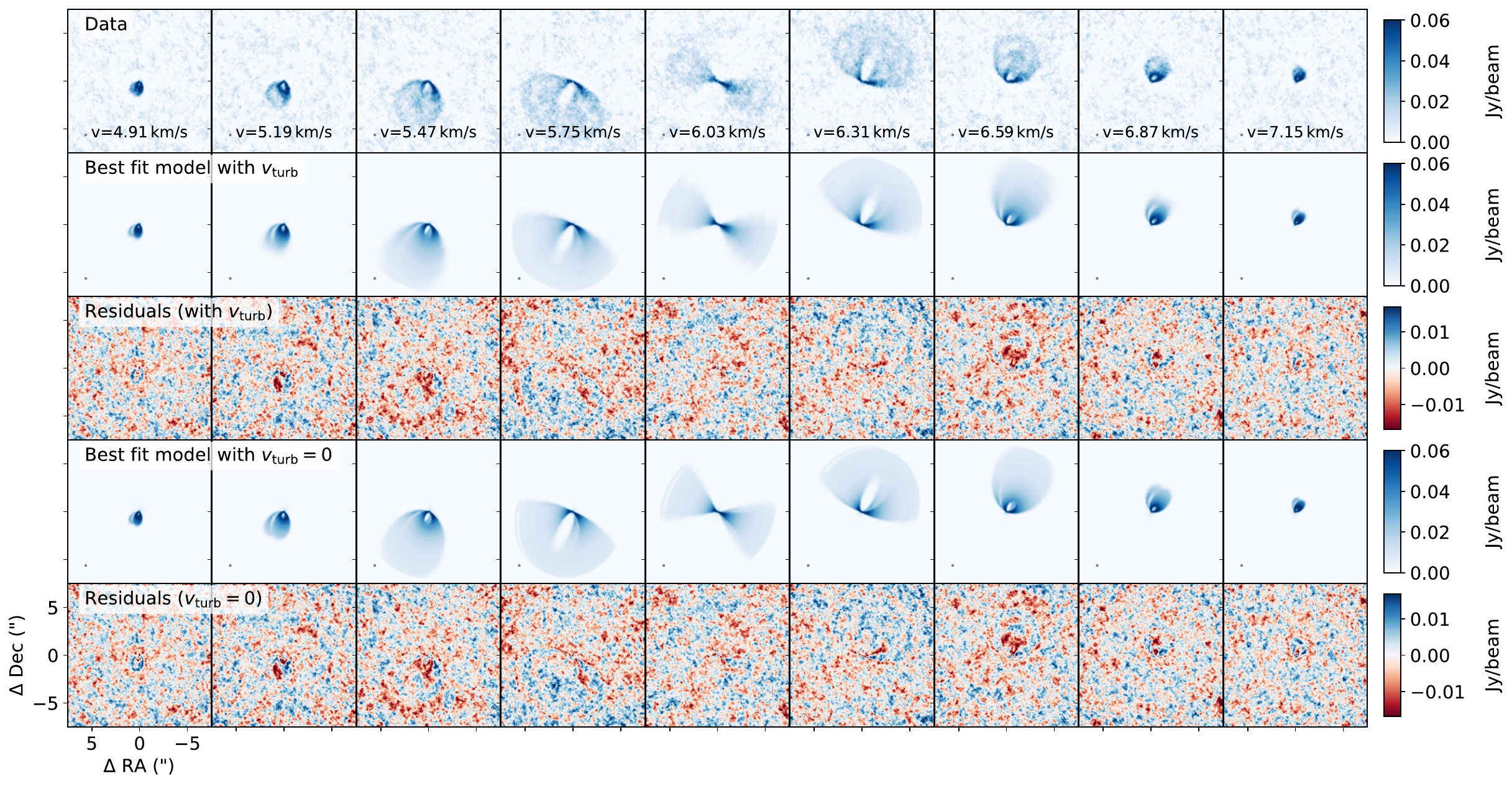}
 \caption{Channel maps of DM Tau in $^{12}$CO $J$=3–2 at 0\farcs15 resolution. The top row shows the observed data, the second row our best-fit model with non-zero turbulence, and the third row the residuals from subtracting this model from the data. The fourth and fifth rows show the model and residuals for the best fit zero-turbulence model for comparison. Each residual color bar spans $\pm3\sigma$ flux for the data cube.}
    \label{fig:image best CO}
\end{figure*}

\begin{figure*}
     \centering
     \begin{subfigure}[b]{0.49\textwidth}
         \centering
         \includegraphics[width=\textwidth]{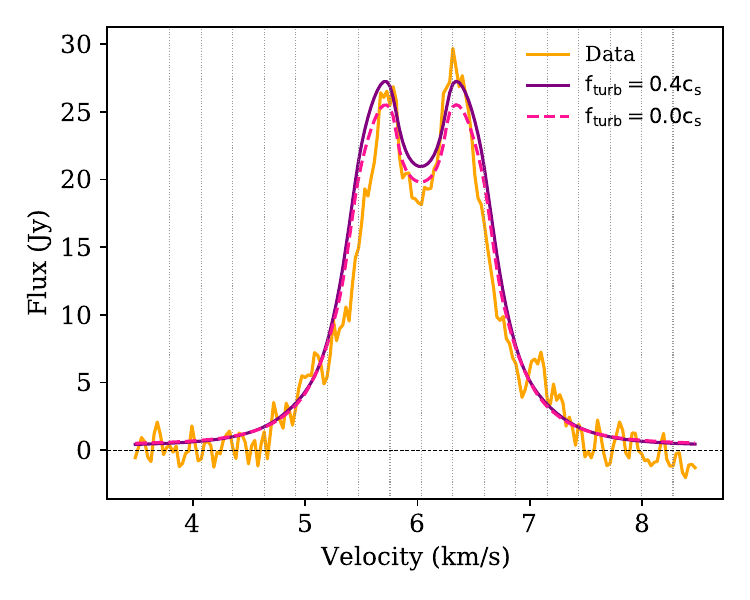}
         \label{fig:co_lineprofile}
     \end{subfigure}
     \begin{subfigure}[b]{0.49\textwidth}
         \centering
         \includegraphics[width=\textwidth]{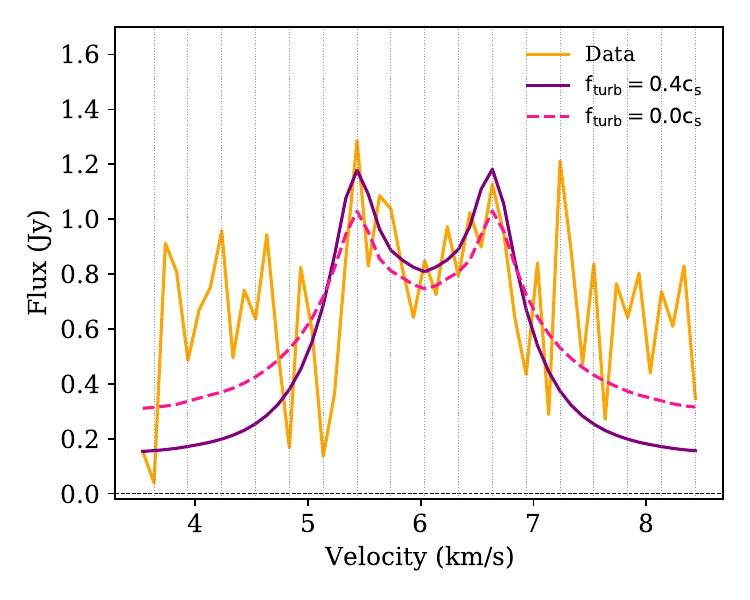}
         \label{fig:cs_lineprofile}
     \end{subfigure}
        \caption{Comparison of observed and modeled integrated line profiles for DM Tau. \textit{Left}: CO $J$=3–2 line profile. \textit{Right}: CS $J$=7–6 line profile. The vertical dotted lines on each panel mark the velocity channels included in the fit. The agreement for CS demonstrates that the CO-derived disk structure can be directly applied to other molecular tracers, allowing differences in emission to be interpreted primarily in terms of chemistry and excitation rather than large-scale structural variations.}
        \label{fig: lineprofiles}
\end{figure*}

\begin{figure*}
	\includegraphics[width=\textwidth]{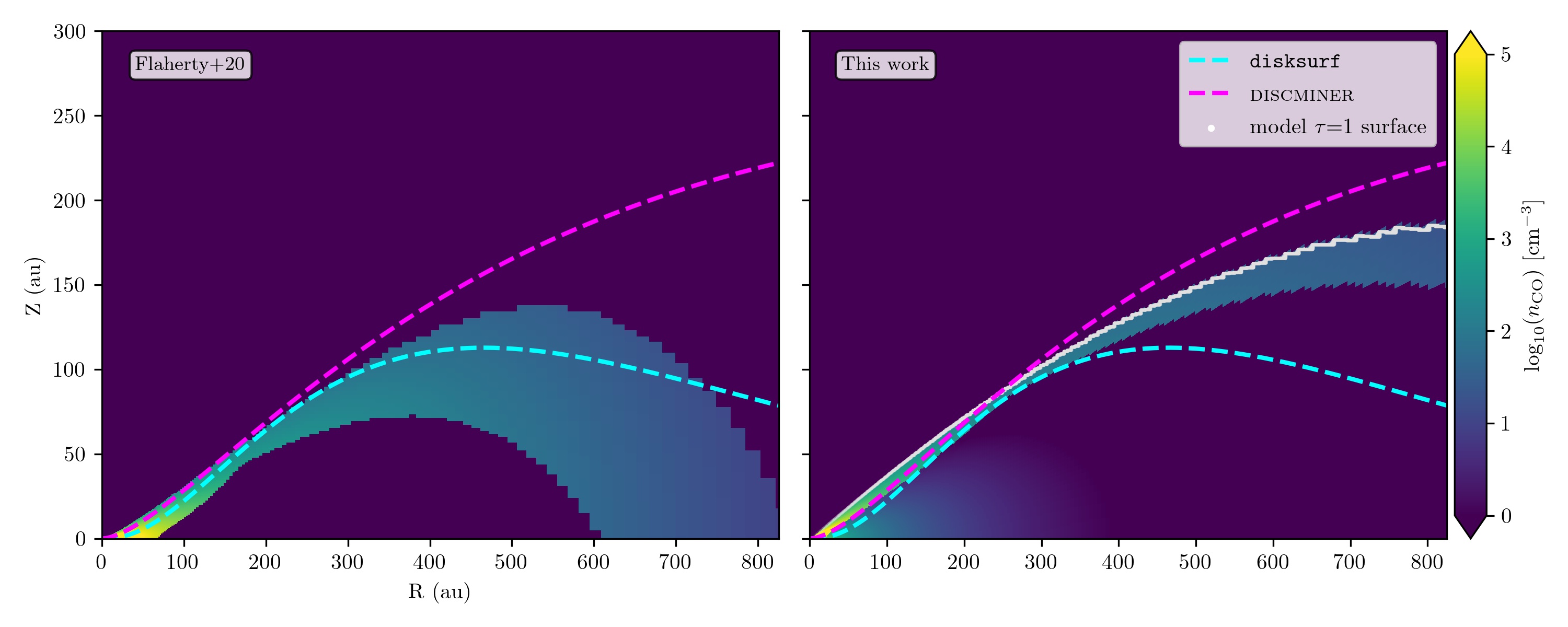}
 \caption{CO number density in DM Tau. The cyan and magenta dashed lines show the exoALMA-derived emission heights from \citet{Galloway-Sprietsma_exoALMA} using \textsc{DiscMiner} and \texttt{disksurf}. \textit{Left}: $J=2-1$ model from \citet{flaherty20}. \textit{Right}: Our $J=3-2$ model.}
    \label{fig:emitting layers co}
\end{figure*}

\begin{figure*}
	\includegraphics[width=\textwidth]{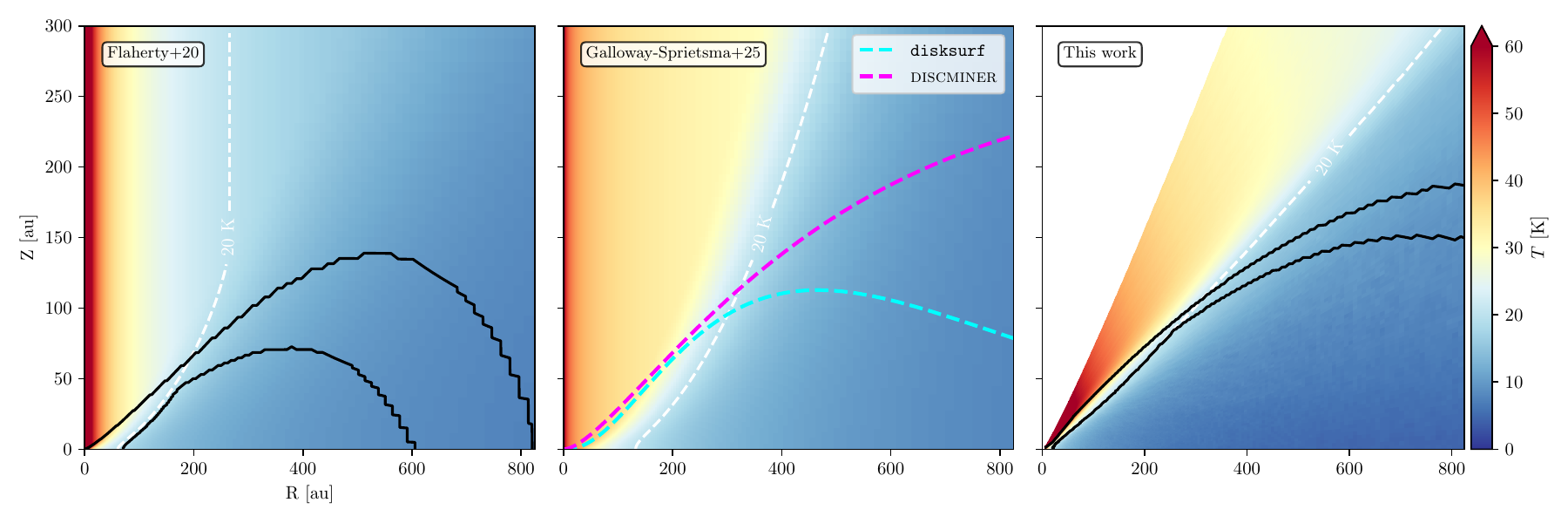}
 \caption{Comparison of temperature structures from different models. Each panel shows the gas temperature as a function of radius and height for (left to right) the models from \citet{flaherty20}, \citet{Galloway-Sprietsma_exoALMA}, and this work. The white contour indicates 20 K, which we used as our CO freeze-out temperature. The black contours trace the height enclosing the top 95\% of the model CO number density at each radius, highlighting the bulk CO while excluding low-density photodesorbed regions. Colored contours indicate the exoALMA emission heights.}
    \label{fig:temp comparison co}
\end{figure*}

We first constrain the disk structure of DM~Tau by fitting the $^{12}$CO $J$=3–2 emission. Figure~\ref{fig:image best CO} shows representative channel maps of our best-fitting model alongside the data and residuals, with the corresponding parameters listed in Table \ref{tab:priors_posteriors}. The model reproduces the overall morphology of the emission across all 17 channels, with the root-mean-square of the residual cube consistent with the image noise, indicating that the model captures the bulk of the emission. The integrated line profile (Figure~\ref{fig: lineprofiles}, left) also shows good agreement with the data. The model slightly overestimates the total flux, with the largest discrepancy of order $\sim$10\% in the systemic channel. This discrepancy arises from small residuals integrated over many pixels: on a per channel basis, the standard deviation of the residuals remain consistent with the image noise (0.0056 Jy/beam), and the mean residual (0.00545 Jy/beam) is below the per-channel rms. The maximum rms in any individual channel is 0.0059 Jy/beam which is also consistent with the image noise. The model reproduces the line wings and overall line shape well.

The posterior distribution (Appendix~\ref{sec: appendix}) shows narrow ranges for the fitted parameters, corresponding to the small uncertainties listed in Table \ref{tab:priors_posteriors}. 
However, these represent only the statistical uncertainties within our adopted model. We neglect pixel correlations in the noise from the imaging process, which can cause underestimation by an order of magnitude \citep{hilder_exoALMA}. Additionally, our smooth and axisymmetric disk model does not completely capture complex substructures in the data \citep{stadler_exoALMA}. Therefore the true uncertainties are larger than the quoted credible intervals.

The posterior mean inclination is $39.7^\circ$ and the position angle is $155.4^\circ$, both in excellent agreement with previous measurements \citep{flaherty20, curone_exoALMA, izquierdo_exoALMA}. The critical radius is $R_c \sim 336$ au with a surface density gradient $\gamma \sim 1.55$, indicating an extended disk with a steep outer density profile. Our best-fit stellar mass is $0.483 \pm 0.004 M_\odot$, consistent with the dynamical mass obtained by \citet{longarini_exoALMA} of $0.468^{+0.014}_{-0.015}$, but slightly lower than the $0.54 M_\odot$ adopted by \citet{flaherty20}. This difference likely reflects variations in modeling assumptions, such as the disk structure and temperature profile, rather than statistical uncertainty.
Overall, this highlights that the derived parameters are sensitive to model framework, and that our results cannot capture uncertainties caused by unmodeled features in the data.

Our model also constrains the global disk structure. The vertical structure is characterized by a scale height $h_0 = 20.6$ au at 150 au and a flaring index $\Psi = 1.16$, implying a moderately flared geometry. 

The relatively large value of $\alpha_{\rm dust}$ results from the fact that fitting procedure for the gas disk is not sensitive to the continuum emission, hence while the model prefers non-settled small (micron-sized) grains it does not imply much about the scale height of millimeter-emitting grains.

\subsection{Turbulent line broadening and disk thermal structure} \label{sec: turbulence results}
A key outcome of the modeling is the constraint on the non-thermal contribution to the linewidth. We measure $f_{\rm turb} = 0.404^{+0.016}_{-0.032} \ c_s$, corresponding to $\alpha~=~0.16^{+0.01}_{-0.02}$ and an average non-thermal broadening of $\sim$180 m~s$^{-1}$ in the CO-emitting region. 
To illustrate the impact of turbulence, we also ran a fit using models with no $f_{\rm turb}$ contribution. While the channel maps of the zero-turbulence model (also shown in Figure \ref{fig:image best CO}) are visually similar to the best-fit turbulent model, the fitting clearly favors a non-zero turbulence value. We also see this in the integrated line profile (Figure \ref{fig: lineprofiles}, left), where the model with no $f_{\rm turb}$ does not fit the peak intensity. This preference is reflected quantitatively in the Akaike Information Criterion \citep{akaike74}, with the free-turbulence model strongly favored over the zero-turbulence model ($\Delta$AIC $\approx 1.25\times10^{4}$), which confirms that including some level of turbulence improves the fit. Our data have a spatial resolution of 0\farcs15 ($\sim$22~AU at the distance of the disk), which resolves the largest disk structures but smears smaller features, and this can partially mask the subtle line broadening introduced by turbulence. Consequently, model fitting, rather than visual inspection alone, is required to robustly detect the non-thermal motions.  
The inferred level from our fit is even higher than the previously determined value of \citet{flaherty20}, who reported 0.25–0.33 $c_s$ from a parametric fit to $^{12}$CO $J$=2–1.

Figure~\ref{fig:emitting layers co} shows the CO number density of our $J=3$–2 model (right) compared with the $J=2$–1 model of \citet{flaherty20} (left). 
We further overplot our model $\tau =1$ surface (shown in white, right) and the CO emission heights derived from the $J=3-2$ exoALMA observations \citep{Galloway-Sprietsma_exoALMA}, obtained using \textsc{DiscMiner} \citep{izquierdo21} and \texttt{disksurf} \citep{disksurf}.
Although our model, and that of \citet{flaherty20}, are based on different transitions, we found a negligible difference between the $\tau=1$ surfaces of the $J=3-2$ and $J = 2-1$ lines in our model. While the underlying CO number density in both models is of the same order of magnitude, the spatial distributions vary, due to differences in modeling assumptions and constraints and the higher spatial and spectral resolution of our exoALMA observations.

The comparison reveals an interesting trend. In the inner disk ($R \sim 250-300$ au), all four measurements --- \citet{flaherty20} in $J=$2-1, our $J$=3-2 model, and the exoALMA emission heights --- are closely aligned, suggesting that the vertical structure is able to be well constrained by a variety of modeling approaches in this region. Beyond $\sim~300$ au, differences become apparent: our 3-2 model and the \textsc{DiscMiner} surfaces reach higher altitudes of 150--200 au, while the \texttt{disksurf} height remains lower, similar to that found by \citet{flaherty20}. This suggests that differences in modeling approach and sensitivity to the diffuse outer disk gas likely drive the elevated surfaces in our model and the \textsc{DiscMiner} emission surface. 

To examine the underlying thermal structure, Figure~\ref{fig:temp comparison co} compares the temperature distribution of our model (right) with those from \citet{flaherty20} and the exoALMA analysis of \citet{Galloway-Sprietsma_exoALMA}, shown on the left and middle panels, respectively. The white dashed contours mark our 20~K freeze-out boundary, while the solid black contours and dashed colored curves show the corresponding emission regions for each analysis. Despite differences in the detailed thermal prescriptions, all models show some regions of temperatures of $\sim$20–30~K within the CO-emitting layer. The emission is dominated by regions where CO remains in the gas phase, above the freeze-out threshold of $\sim 20$ K, typically within $ R\sim 200-300$~au. As a result, the local thermal conditions in the emitting layer set the observed CO emission, making it relatively insensitive to the details of the global temperature parameterization.

\subsection{CS fitting}\label{sec:cs results}

\begin{figure*}
\includegraphics[width=\textwidth]{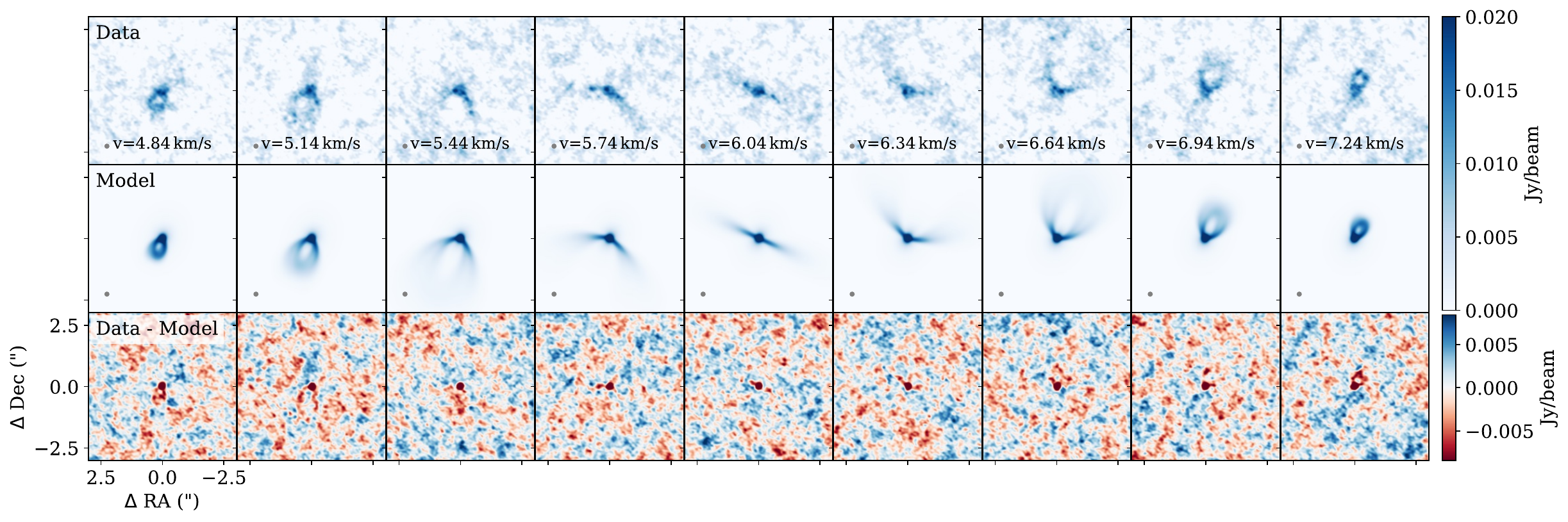}
 \caption{Channel maps of DM~Tau in CS \textit{J}=7--6 emission at 0\farcs15 resolution. The top row shows the observed data, the middle row the equivalent channels for our best-fit model and the bottom row the residuals from subtracting this model from the data. The residual color bar represents $\pm3\sigma$ flux of the data.}
    \label{fig:CS fit}
\end{figure*}

\begin{figure*}
	\includegraphics[width=\textwidth]{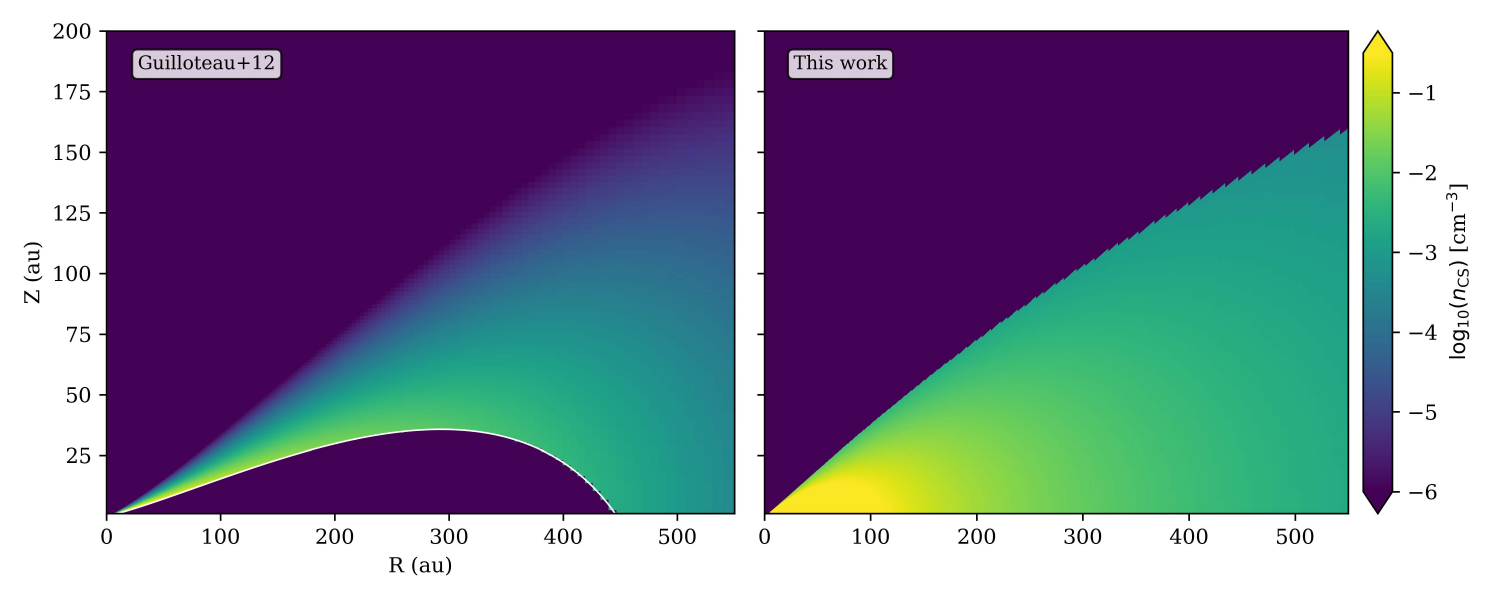}
 \caption{CS number density in DM Tau. \textit{Left}: CS $J$=3–2 model from \citet{guilloteau12}, which uses a radially decreasing abundance and a hydrogen column density depletion threshold of $10^{21.7} \ \rm cm^{-2}$ (white curve). \textit{Right}: Our CS $J$=7–6 model, adopting a constant abundance $\rm log10(X_{\rm CS}) = -9.4$ and a freeze-out threshold of 30~K.}
    \label{fig:emitting layers cs}
\end{figure*}

We verify the robustness of the CO-derived disk structure by using it as a base model for the CS $J$=7–6 emission. The best-fit model (parameters listed in Table~\ref{tab:priors_posteriors}) reproduces both the channel maps (Figure~\ref{fig:CS fit}) and the integrated line profile (Figure~\ref{fig: lineprofiles}, right), with residuals at the noise level. The integrated line profile for CS is significantly noisier than that of CO due to its lower abundance and weaker emission, which reduces the signal-to-noise. Similar to our CO comparison, we also plot the spectrum of the best fitting model with no turbulent contribution (Figure \ref{fig: lineprofiles}, right) and once again see that a significant level of non-thermal broadening in the model is required to reproduce the peak emission in the data. We also observe a consistent over-prediction of emission near the disk center in all channels (Figure~\ref{fig:CS fit}), producing a central residual. This likely results from the continuum emission not being modeled accurately --- an effect that is negligible for CO but becomes more significant for the weaker CS emission ----- rather than a fundamental issue with the underlying CO-based disk model. Therefore we only trust the model at radii larger than the extent of the continuum emission ($R \gtrsim  119$~au).

The inferred CS abundance of $\log_{10}(X_{\rm CS}) \approx -9.4$ corresponds to a CS/CO abundance ratio of order $10^{-5}$. The freeze-out fraction for CS is $f_{\rm remain} \sim 0.997$, indicating that essentially all CS remains in the gas phase, with freeze-out playing a negligible role. In contrast, CO is almost entirely frozen onto grains at temperatures below 20~K ($f_{\rm remain} \sim 0.002$). The CS emission can therefore be reproduced without invoking significant depletion onto grains, and without altering the underlying disk structure derived from CO. This again confirms that the CO-based model provides a reliable framework for interpreting multiple molecular tracers, with chemical differences arising primarily from abundance variations rather than structural uncertainties.

We can also compare our resulting CS model with that of \citet{guilloteau12}, who fit the CS $J=3-2$ transition using a similar density parameterization but a radially decreasing CS abundance ($4.2\times10^{-10} \times (r/300~\mathrm{AU})^{-0.39}$) and a column density threshold that removes CS from the midplane. Figure \ref{fig:emitting layers cs} shows the comparison of the resulting column densities. Because \citet{guilloteau12} adopted a decreasing abundance profile, our constant abundance of $X_{\rm CS} = 10^{-9.4}$ yields number densities that are on average similar across the disk, though slightly lower in the inner regions and somewhat higher in the outer disk. To reproduce the observed $J=7-6$ emission, our model allows CS to remain in the gas phase throughout the disk, effectively turning off freeze-out. This behavior is consistent with the findings of \citet{guilloteau12}, who showed that CS can exist at temperatures well below its evaporation temperature ($\sim$30~K). In their study, the vertical extent of the molecular-rich layer was primarily regulated by UV penetration, photodesorption, and photodissociation rather than temperature, with thermal desorption becoming significant only above $\sim$30~K.

Overall, these results show that the CS $J=7-6$ emission can be reproduced within the CO-derived disk framework, with differences from the \citet{guilloteau12} $J=3-2$ model primarily driven by our lower, constant abundance and the absence of a tuneable column density threshold when reproducing photodesorption effects. The persistence of CS down to the midplane indicates that it is not in thermo-chemical equilibrium, emphasizing the role of chemical processes and UV-driven effects in shaping its vertical distribution, which we will discuss more in Section \ref{sec: discussion cs}.

\section{Discussion}\label{sec:discussion}

\subsection{Turbulence in the DM~Tau disk}\label{sec: discussion turbulence}

Our modeling of the $^{12}$CO $J$=3–2 emission in DM~Tau constrains the non-thermal contribution to the linewidth to $f_{\rm turb} = 0.403^{+0.008}_{-0.012}\ c_s$, corresponding to an average non-thermal broadening of $\sim$180~m~s$^{-1}$ in the CO-emitting layers. This level of turbulence is significantly higher than previous estimates from the CO $J$=2–1 transition (\citealt{flaherty20}, $0.25$–$0.33\ c_s$) and inconsistent with a completely laminar disk.

In our model, the non-thermal broadening is parameterized as a fixed fraction of the local sound speed, implying a spatially constant Mach number throughout the emitting region. This approach follows other recent analyses \citep[e.g.][]{flaherty20} and provides a convenient way to compare systems, though it likely oversimplifies the true structure of turbulence in disks. Studies that adopt radially varying or non-parametric prescriptions \citep{guilloteau12, teague16} and numerical simulations \citep{simon15, barraza-alfaro_exoALMA} suggest that turbulence generally increases with height above the midplane as instabilities such as the VSI or MRI become more effective. Consequently, our single-parameter description should be viewed as an average measure of the turbulent amplitude within the CO-emitting layers rather than a detailed vertical profile.

The elevated turbulence inferred here may also reflect additional physical and observational effects. Higher-$J$ CO transitions probe warmer, more elevated layers of the disk which are naturally more dynamic and therefore more susceptible to the kinds of motions driven by the instabilities discussed above. As mentioned in Section \ref{sec: turbulence results}, we do not see a notable difference between the $J=3-2$ and $2-1$ emission surfaces for our models, but compared to the \citet{flaherty20} model (left panel of Figure \ref{fig:emitting layers co}) our surface in the outer disk is higher, which could be leading to our higher inferred value of $f_{\rm turb}$.
If the disk were fully turbulent, we would expect to observe large scale deviations from Keplerian rotation at the exoALMA spectral resolution as demonstrated by \citet{barraza-alfaro_exoALMA}, but our data do not show such global motions. Instead, localized small-scale motions, potentially induced by embedded planets or local instabilities, may contribute to the line broadening: \citet{pinte_exoALMA} identified a potential velocity kink and filamentary structures between the upper and lower disk surfaces in DM~Tau which result in subtle deviations from pure Keplerian rotation, the same as those we can identify in our residuals (see Section \ref{sec: discussion substructure}. While our high-resolution exoALMA data of DM~Tau do not show large-scale non-Keplerian arcs such as those seen around CQ~Tau, HD~135344B, J1604 and MWC~758, these localized kinematic features in DM~Tau could account for part of the measured non-thermal linewidth.  Taken together, the inferred non-thermal broadening and the absence of global kinematic deviations imply that the observed line widths could reflect localized or small-scale motions, though distinguishing these from genuinely turbulent processes remains challenging.

An important consideration when comparing our gas temperature and turbulence measurements to other studies is the potential effect of amplitude calibration uncertainties. Previous analyses, which derived gas temperatures directly from the observed CO line intensities, such as \citet{flaherty20}, are susceptible to systematic errors when the flux scale is translated directly into temperature uncertainties. This in turn affects the inferred thermal and non-thermal broadening. Our radiative transfer approach is fundamentally different, and while we do not fit explicitly for temperature, flux calibration errors could in principle affect our inferred disk properties.

We performed parameter studies and found that the parameters which dominate the observed line emission in our models are the stellar radius (which sets the luminosity and therefore the temperature structure), the external radiation field (which provides additional heating in the upper disk layers) and the CO abundance (which converts the temperature and density structure into observable emission). We fit all of these parameters simultaneously, and our posterior distributions (Figure \ref{fig:co corner}) show that they are well-constrained without significant degeneracies with each other or with the turbulent broadening, indicating that systematic flux calibration errors are not being absorbed into biases in individual parameters. The variance inflation parameter $\ln q$ further mitigates the impact of systematic flux scale mismatches.

Crucially, our turbulence measurement is constrained primarily by the local linewidths rather than the absolute flux level. Since flux calibration errors do not translate directly to temperature uncertainties in our model unlike parametric fits to the data, our turbulence measurement is therefore relatively insensitive to systematic uncertainties in the flux calibration. We therefore expect our turbulence constraints to be robust against amplitude calibration uncertainties at the $\sim$10–15\% level typical of ALMA observations.

\subsection{Substructure in the DM~Tau disk}\label{sec: discussion substructure}

\begin{figure*}
	\includegraphics[width=\textwidth]{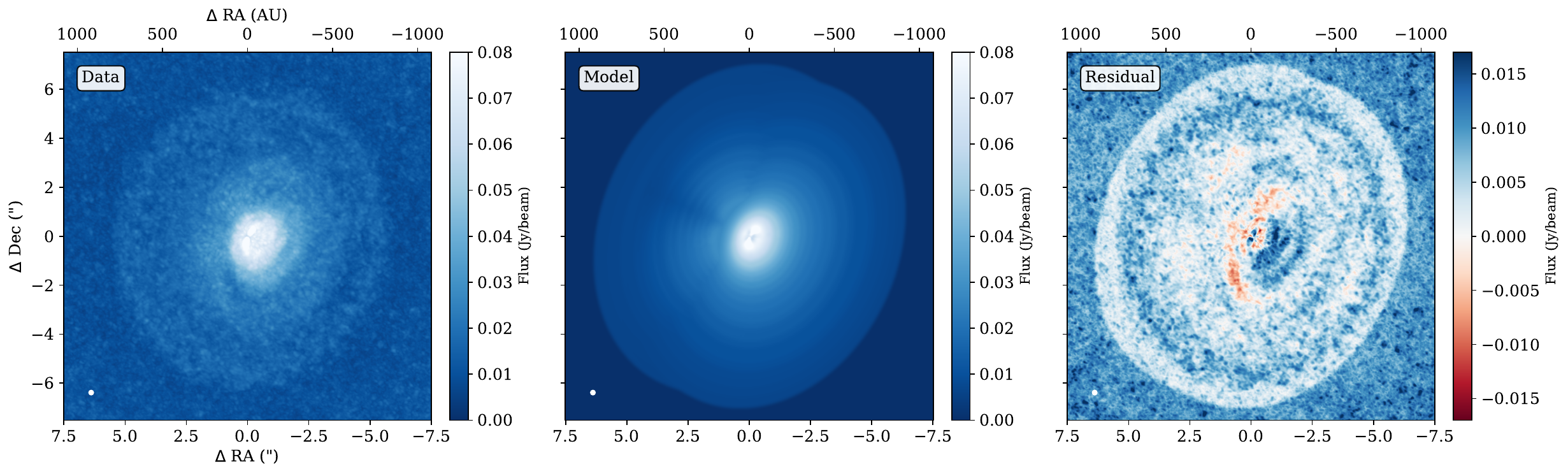}
    \caption{$^{12}$CO peak intensity maps. \textit{Left}: observed data. \textit{Middle}: best-fit model. \textit{Right}: residual map (data minus model), highlighting regions where the model over- or under-predicts the emission. 
    .}
   \label{fig:residual_8_co}
\end{figure*}

\begin{figure*}
	\includegraphics[width=\textwidth]{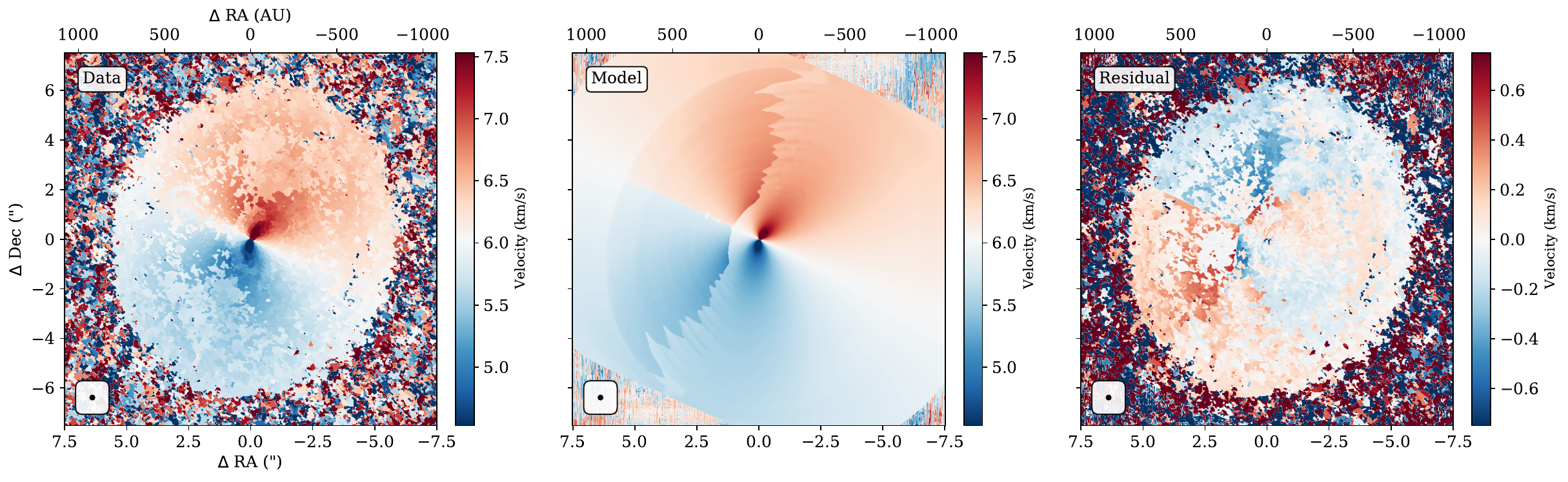}
    \caption{$^{12}$CO peak velocity maps. \textit{Left}: observed data. \textit{Middle}: best-fit model. \textit{Right}: residual map (data minus model), showing differences in the velocity of peak emission across the disk. 
    }
   \label{fig:residual_9_co}
\end{figure*}



As mentioned in Section \ref{sec:model}, we have assumed a smooth disk for DM Tau as including substructure in the model would drastically widen the parameter space.
However, the disk around DM Tau was referred to as a ``proto-asteroid belt" by \citet{kudo18}, with ALMA observations of the dust continuum emission revealing that the disk is comprised of multiple dust rings and an inner disk \citep{hashimoto21,curone_exoALMA}. These substructures could reflect the kinematics of the gas in the outer disk, although our modeling does not attempt to capture such effects.

As a result of our smooth model assumption, residual moment maps of the $^{12}$CO emission reveal subtle deviations which may provide clues as to the localized dynamical processes in DM~Tau. Figures \ref{fig:residual_8_co} and \ref{fig:residual_9_co} shows examples of these residuals in the peak intensity and velocity maps respectively, where the model over- or under-predicts the emission or the local line-of-sight velocity in the inner disk. While the origin of the substructures is unclear in these residual maps, the spiral-like morphology may be caused by embedded planets. An outer feature at $\sim$600~au (4\farcs2, the dark blue ring) appears in both moments; its morphology is more consistent with the separation between the front and back CO-emitting surfaces rather than a true over-density. Across the disk, the amplitude of these CO residuals reaches 10–20\% of the peak intensity, which demonstrates that even minor deviations from a smooth structure are detectable. 


The strength of our approach is in the use of full radiative transfer modelling, which self-consistently constrains the disk's vertical and thermal structure. While other studies of the exoALMA disks \citep[e.g.][]{izquierdo_exoALMA, Galloway-Sprietsma_exoALMA, barraza-alfaro_exoALMA} adopt smooth parametric models as their baseline, our radiative transfer framework provides a physically consistent treatment of temperature, density and emission. This provides us with a more realistic baseline for identifying residuals from our smooth models and allows subtle dynamical or chemical substructures associated with ongoing planet formation to be identified with greater confidence.

\subsection{What is causing the CS emission?} \label{sec: discussion cs}

We find that significant CS emission arises from regions of the disk below 30~K, where CS would normally be expected to freeze onto dust grains. In our model, photodesorption is included but is not strong enough to reproduce the observed $J=7$–6 emission on its own. Instead, the emission requires that CS remain largely in the gas phase throughout the disk, implying that freeze-out is strongly suppressed or that additional non-equilibrium processes act to replenish gas-phase CS in cold regions.

Physically, the $f_{\rm remain}$ parameter may represent the cumulative effect of processes that prevent complete CS depletion in the disk midplane. These could include cosmic-ray or chemical desorption and vertical mixing that transports CS-rich gas from warmer layers. Other mechanisms could also contribute to the formation of CS at low temperatures. For example, shocks caused by planet wakes can destroy CO, liberating atomic carbon which then reacts with available sulfur to form CS \citep{law25}. A similar process occurs in AGB winds, where CS forms after C is liberated from CO through shocks \citep{Cherchneff2006,Danilovich2019}, especially when there is little free C available otherwise.

Comparing our distribution to theoretical chemical models, \citet{Walsh2010} found that including photodesorption increases CS abundances to values comparable to those in our model (i.e. $\rm log10(X_{CS}) \sim -9$). However, in their model the abundance varies with radius and height, tapering off towards the midplane, while our model assumes a constant abundance and negligible freeze-out. This difference may indicate that our simplified prescription effectively compensates for missing non-thermal desorption or mixing processes that maintain CS in the gas phase. A more realistic treatment would likely allow the degree of depletion to vary with both radius and height, since the stellar parameters adopted by \citet{Walsh2010} are similar to those used here for DM~Tau.

To observationally constrain the dominant mechanisms responsible for sustaining CS, it will be important to study CS emission across a larger sample of disks spanning a range of stellar and disk conditions, including systems with evidence for planetary companions, to determine how the CS distribution varies with disk structure, irradiation environment, and planet–disk interactions.

\section{Conclusions}
\label{sec:conclusions}

Leveraging the high-resolution $^{12}$CO \textit{J}=3–2 and CS \textit{J}=7–6 data of the DM Tau disk from the exoALMA survey, we perform full radiative transfer modeling with \textsc{mcfost} to constrain the disk structure and non-thermal line broadening. We find

\begin{enumerate}
\item Our best-fit CO model reproduces both the channel maps and the integrated line profile, yielding a measured non-thermal broadening of $f_{\rm turb}~=~0.4~c_{\rm s}$, slightly higher than previously reported.

\item We find that the CS number density in the midplane is very similar to previous models, but due to our model parameterization (using a constant CS abundance and allowing freeze-out to be supressed) produces emission that extends more uniformly and deeper into the disk. Future work could incorporate a more physically motivated treatment of freeze-out and desorption to better capture the underlying chemistry.

\item  Residual maps of $^{12}$CO (Figure~\ref{fig:residual_8_co}) reveal subtle deviations from our smooth disk model, indicating the presence of weak substructures that may trace dynamical perturbations or variations in local chemistry within the DM~Tau disk. 
\end{enumerate}

By combining high-resolution data from surveys like exoALMA with detailed radiative transfer modeling, it is now possible to better characterize disk turbulence, vertical structure, molecular distributions, and low-level substructures across multiple sources. Applying this approach to other disks would help to reveal trends in disk physics and chemistry and place individual systems such as DM Tau in the broader context of protoplanetary disk evolution and planet formation.

\section*{Acknowledgments}

Caitlyn Hardiman (CH) and TH are funded by Research Training Program Scholarships from the Australian Government. CH, DJP, CP, TH, and IH acknowledge funding from the Australian Research Council via DP220103767 and DP240103290. CH acknowledges support from the Astronomical Society of Australia. 
TD is supported in part by the Australian Research Council through a Discovery Early Career Researcher Award (DE230100183).
GR acknowledges funding from the Fondazione Cariplo, grant no. 2022-1217, and the European Research Council (ERC) under the European Union's Horizon Europe Research \& Innovation Programme under grant agreement no. 101039651 (DiscEvol). 
JB acknowledges support from NASA XRP grant No. 80NSSC23K1312. 
MB, DF, JS, and IH have received funding from the European Research Council (ERC) under the European Union's Horizon 2020 research and innovation programme (PROTOPLANETS, grant agreement No. 101002188). 
NC has received funding from the European Research Council (ERC) under the European Union Horizon Europe research and innovation program (grant agreement No. 101042275, project Stellar-MADE).
PC acknowledges support by the ANID BASAL project FB210003. 
SF is funded by the European Union (ERC, UNVEIL, 101076613), and acknowledges financial contribution from PRIN-MUR 2022YP5ACE. 
MFl has received funding from the European Research Council (ERC) under the European Unions Horizon 2020 research and innovation program (grant agreement No. 757957).
MFu is supported by a Grant-in-Aid from the Japan Society for the Promotion of Science (KAKENHI: No. JP22H01274). 
Cassandra Hall gratefully acknowledges support from the U.S. National Science Foundation Grants 2511673 and 2407679, NRAO SOSPADA-036, National Geographic Society, and the Georgia Museum of Natural History.
JDI acknowledges support from an STFC Ernest Rutherford Fellowship (ST/W004119/1) and a University Academic Fellowship from the University of Leeds. 
Support for AFI was provided by NASA through the NASA Hubble Fellowship grant No. HST-HF2-51532.001-A awarded by the Space Telescope Science Institute, which is operated by the Association of Universities for Research in Astronomy, Inc., for NASA, under contract NAS5-26555. 
GLe and GWF have received funding from the European Research Council (ERC) under the European Union Horizon 2020 research and innovation program (Grant agreement no. 815559 (MHDiscs)). GWF was granted access to the HPC resources of IDRIS under the allocation A0120402231 made by GENCI. 
CL and GLo have received funding from the European Union's Horizon 2020 research and innovation program under the Marie Sklodowska-Curie grant agreement No. 823823 (DUSTBUSTERS). CL acknowledges support from the UK Science and Technology research Council (STFC) via the consolidated grant ST/W000997/1. 
FMe acknowledges funding from the European Research Council (ERC) under the European Union's Horizon Europe research and innovation program (grant agreement No. 101053020, project Dust2Planets).
H-WY acknowledges support from National Science and Technology Council (NSTC) in Taiwan through grant NSTC 113-2112-M-001-035- and from the Academia Sinica Career Development Award (AS-CDA-111-M03). 
TCY is supported by Grant-in-Aid for JSPS Fellows JP23KJ1008. 
AJW has been supported by the European Union’s Horizon 2020 research and innovation programme (Marie Sklodowska-Curie grant agreement No 101104656 and by the Royal Society through a University Research Fellowship, grant number URF\textbackslash R1\textbackslash 241791.
Support for BZ was provided by The Brinson Foundation. 
Views and opinions expressed by ERC-funded scientists are however those of the author(s) only and do not necessarily reflect those of the European Union or the European Research Council. Neither the European Union nor the granting authority can be held responsible for them.
This paper makes use of the following ALMA data: ADS/JAO.ALMA\#2021.1.01123.L. ALMA is a partnership of ESO (representing its member states), NSF (USA) and NINS (Japan), together with NRC (Canada), MOST and ASIAA (Taiwan), and KASI (Republic of Korea), in cooperation with the Republic of Chile. The Joint ALMA Observatory is operated by ESO, AUI/NRAO and NAOJ. 
This work was performed on the OzSTAR national facility at Swinburne University of Technology. The OzSTAR program receives funding in part from the Astronomy National Collaborative Research Infrastructure Strategy (NCRIS) allocation provided by the Australian Government, and from the Victorian Higher Education State Investment Fund (VHESIF) provided by the Victorian Government.

\software{bettermoments \citep{bettermoments}, Matplotlib \citep{Hunter2007}, NumPy \citep{Harris2020}, SciPy \citep{Virtanen2020} and Astropy \citep{astropy2013,astropy2018, astropy2022}.
          }

\appendix

\section{Posterior results}\label{sec: appendix}

\begin{figure*}
	\includegraphics[width=\textwidth]{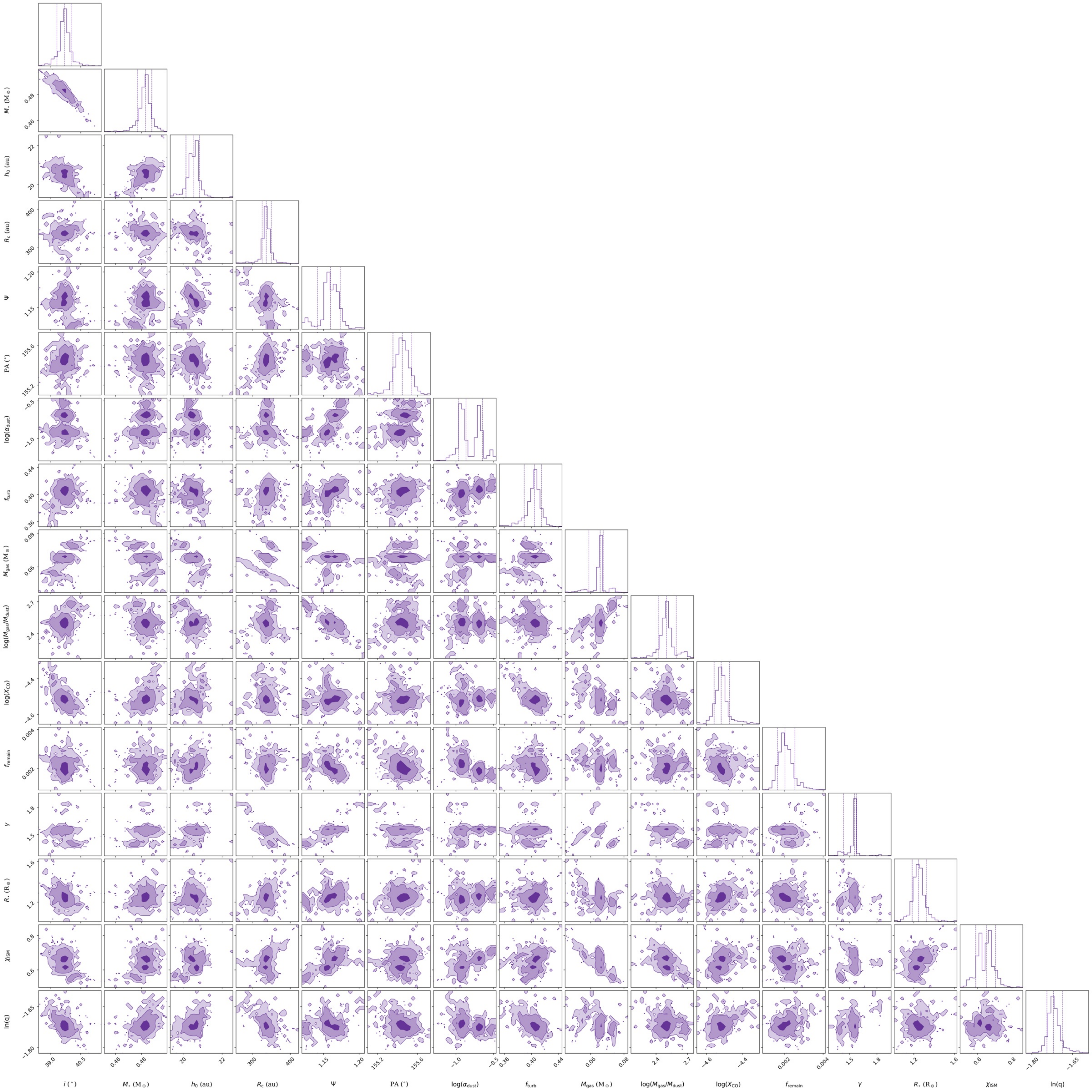}
    \caption{Marginalized one- and two-dimensional posterior distributions of the CO emission fit. Vertical lines in the one-dimensional distributions mark the 10th, 50th, and 90th percentiles.}
    \label{fig:co corner}
\end{figure*}

\begin{figure*}
	\includegraphics[width=\textwidth]{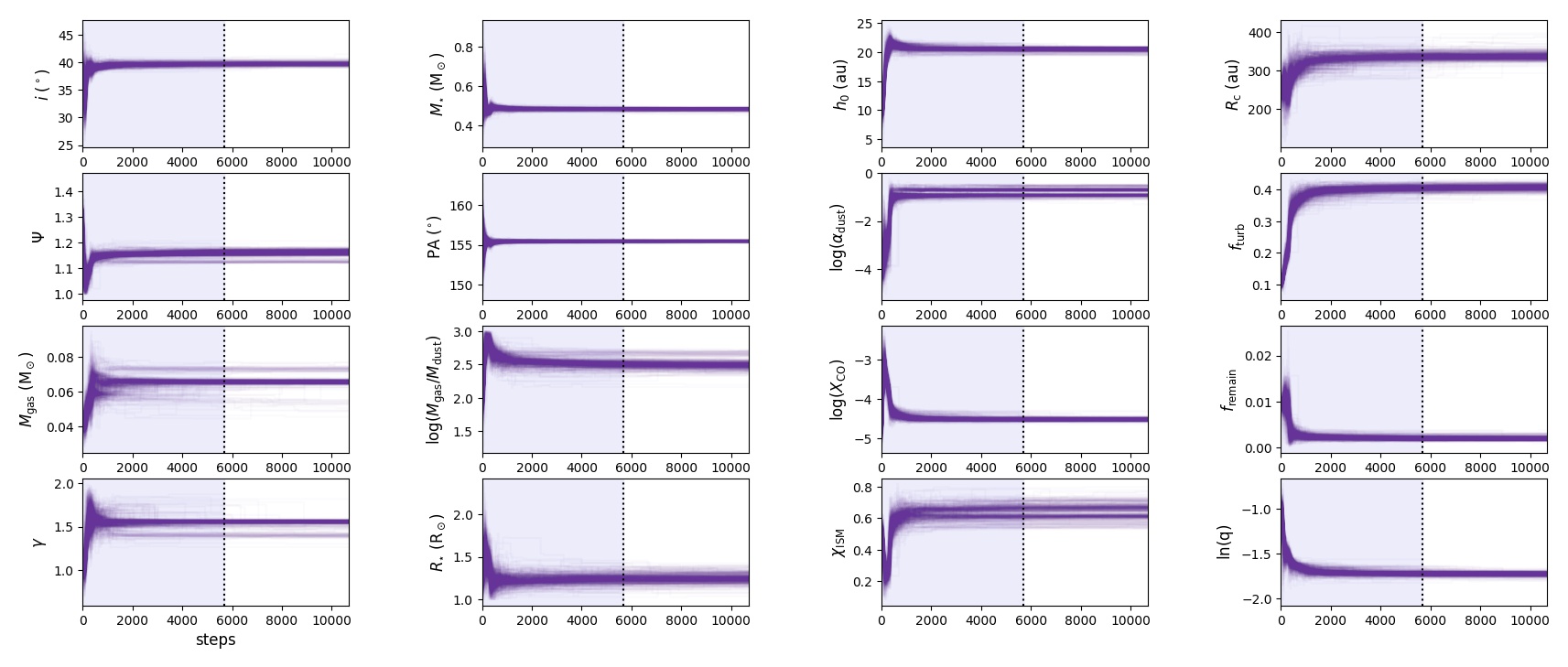}
    \caption{Walker progression for the fit to the CO emission. The vertical dotted line indicates the transition between the burn-in and sampling phases; the burn-in region (shaded in purple) is discarded prior to calculating median parameter values.}
    \label{fig:co traces}
\end{figure*}

Here we present some additional plots showing details of the MCMC fits.
Figure~\ref{fig:co corner} shows our 1- and 2-dimensional posteriors, and Figure~\ref{fig:co traces} shows the walker progression plots for our fit to the CO emission. See Section~\ref{sec:co results} for discussion of these results.

\begin{figure}
\centering
	\includegraphics[width=0.5\columnwidth]{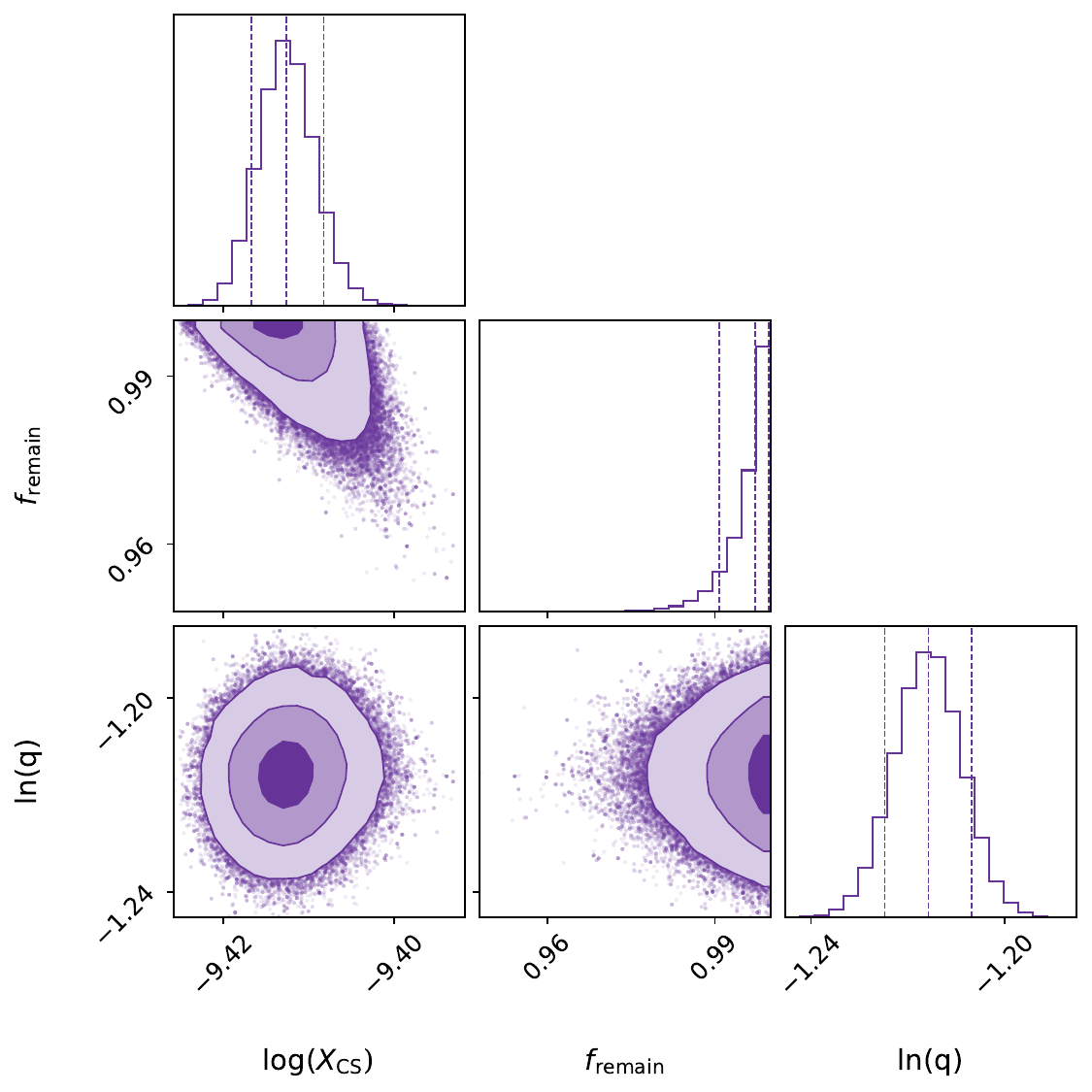}
    \caption{Marginalized one- and two-dimensional posterior distributions of the CS emission fit. Vertical lines in the one-dimensional distributions mark the 10th, 50th, and 90th percentiles.}
    \label{fig:cs corner}
\end{figure}

\begin{figure*}
	\includegraphics[width=\textwidth]{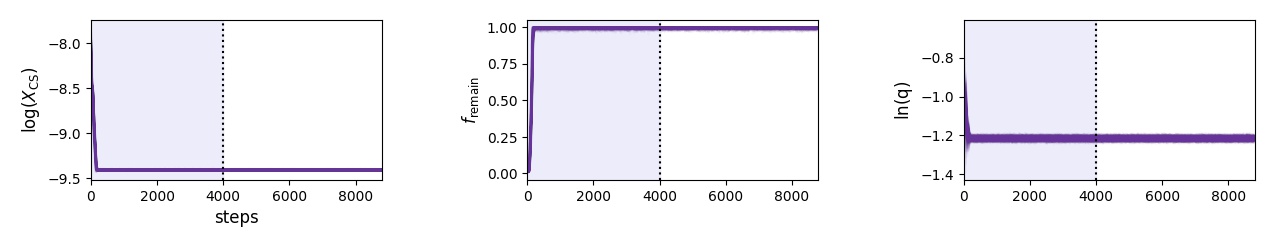}
    \caption{Same as Figure \ref{fig:co traces} but for the fit to the CS emission.}
    \label{fig:cs traces}
\end{figure*}

Figure~\ref{fig:cs corner} shows our 1- and 2-dimensional posteriors, and Figure~\ref{fig:cs traces} shows the walker progression plots for our fit to the CS emission. See Section~\ref{sec:cs results} for discussion of these results.

\bibliography{cait_lib.bib}{}
\bibliographystyle{aasjournal}

\end{document}